\magnification 1200
\baselineskip = 15pt

\font\BLarge=cmbx12
\font\Frak=eufm10
\def\frak#1{{\hbox{\Frak#1}}}

\font\Bbb=msbm10 
\def\BBB#1{\hbox{\Bbb#1}}

\headline={
\hfill 
}

\def\g{{\frak g}}
\def\h{{\frak h}}
\def\s{{\frak s}}
\def\C{\BBB C}
\def\Z{\BBB Z}
\def\N{\BBB N}
\def\tg{\tilde \g}
\def\dg{\dot \g}
\def\D{{\cal D}}
\def\t{{\bf t}}
\def\r{{\bf r}}
\def\m{{\bf m}}
\def\q{{\bf q}}
\def\exp#1{{\rm exp} \bigl( #1 \bigr)}
\def\deg{{\rm deg}}
\def\hgt{{\rm ht}}
\def\mod{{\rm mod}}
\def\ol{\overline}
\def\phi{\varphi}
\def\d{\partial}
\def\dz{D_z}
\def\End{{\rm End}}
\def\zba{\bigl({z_2\over z_1}\bigr)}
\def\zab{\bigl({z_1\over z_2}\bigr)}
\def\K{{\cal K}}

\

\

\

\centerline{\BLarge Principal Vertex Operator Representations} 
\centerline{\BLarge For Toroidal Lie Algebras}

\

\centerline{{\bf Yuly Billig}
\footnote{*}{Work supported by the Natural Sciences and 
Engineering Research Council of Canada.}}

\

\centerline{Department of Mathematics and Statistics,
University of New Brunswick,}
\centerline{Fredericton, N.B., E3B 5A3, Canada} 

\

\

\

{\bf 0. Introduction.}

\

Vertex operators discovered by physicists in string theory have
turned out to be important objects in mathematics. One can
use vertex operators to construct various realizations of the 
irreducible highest weight representations for affine Kac-Moody
algebras. Two of these, the principal and homogeneous realizations,
are of particular interest. The principal vertex operator 
construction for the affine algebra $A_1^{(1)}$ allows one to construct
soliton solutions of the Korteweg~-~de~Vries hierarchy of partial
differential equations. On the other hand, the homogeneous realization 
is linked to the fundamental nonlinear Schr\"odinger equation [Kac].

S.Eswara~Rao and R.V.~Moody in [EM] studied the homogeneous vertex
operator construction for toroidal Lie algebras. The present
paper is devoted to the principal realization.
Here we construct the principal vertex operator representation
for the toroidal Lie algebra $\hat\g$ which is a universal
central extension of $\tg = \dg \otimes \C [t_0^\pm,\ldots, t_n^\pm]$,
in which $\dg$ is a simply-laced simple finite-dimensional Lie algebra
over $\C$. This generalizes the principal vertex operator realization
of the basic representations of affine Lie algebras constructed in [KKLW].

 We add the Lie algebra $\D$ of vector fields on a torus 
to the toroidal Lie algebra $\hat\g$ to form a larger algebra $\g$.
This is necessary in order to have a sufficiently large principal
Heisenberg subalgebra. To construct a representation of $\g$ we 
consider the standard representation of the principal Heisenberg
subalgebra on the Fock space $F$ and then extend it to all of $\g$
by means of the vertex operators. The module $F$ is irreducible as
a module over $\g$, and it is reducible over $\hat\g$.
It will be interesting to see how this representation fits
in the framework of the Verma modules over the toroidal Lie algebras
studied in [BC].

 The vertex operator representations constructed
in this paper for the toroidal Lie algebras may also be
useful for quantum field theories in space-time
of more than two dimensions [IKUX].

In section 1 we recall the construction of the toroidal Lie algebra
and in section 2 we sketch the principal vertex operator realization
of the basic representations of affine Lie algebras. In section 3
we present the principal Heisenberg subalgebra of $\g$ and define
its action on the Fock space. We also introduce there vertex operators
that will allow us to extend this action to $\g$. At the end of the section
we state the main theorem of the paper. The proof of the main theorem
occupies section 4. In the final section we construct the analogues of 
the Sugawara operators.  

\

\

{\bf 1. Definitions and notations.}

\

Let $\dot \g$ be a simple finite-dimensional complex Lie
algebra of type $A_\ell, D_\ell$ or $E_\ell$
 (i.e., simply-laced). The Killing form $( \cdot | \cdot )$
is non-degenerate on the Cartan subalgebra $\dot \h$ of $\dg$
and induces the map $\nu : \dot\h \rightarrow \dot\h^*$.
Let $\dot \Delta$ be the root system of $\dot \g$.
Since $\dot \g$ is simply-laced then $(\alpha | \alpha) 
= 2$ for all nonzero roots $\alpha \in \dot \Delta$.
and for root elements $e_\alpha \in \dot\g^\alpha, 
e_{-\alpha} \in \dot\g^{-\alpha} $ we have $[e_\alpha,e_{-\alpha}] = 
(e_\alpha | e_{-\alpha}) \nu^{-1}(\alpha)$.

To construct a toroidal algebra let us
consider a tensor product of
$\dot \g$ with the algebra of Laurent polynomials
in $n + 1$ variable:
$$ \tilde \g = \dot \g \otimes \C [t_0^\pm,
t_1^\pm,\ldots, t_n^\pm] .$$
Then we define the toroidal Lie algebra corresponding to $\dg$ as
the universal central extension of $\tg$.
The explicit construction of this extension is known from the works 
[Kas], [MEY]. Let $\dot \K$ be an 
$(n + 1)$-dimensional space with the basis
$\{ K_0, K_1,\ldots, K_n\}$. Consider a derivation 
$d_p = t_p {d\over dt_p}$ of the algebra 
$\C [t_0^\pm,t_1^\pm,\ldots, t_n^\pm]$ which can be
naturally extended on the tensor product
$$\tilde \K = \dot \K \otimes \C [t_0^\pm,
t_1^\pm,\ldots, t_n^\pm] .$$

The space of the universal central extension
of $\tg$ is 
$$ \K = \tilde \K / d\tilde \K ,$$
where 
$$d\tilde \K = \bigl\{ \sum\limits_{p = 0}^n
K_p \otimes d_p(f) \big| f \in \C [t_0^\pm,
t_1^\pm,\ldots, t_n^\pm]  \bigr\} \subset \tilde \K .$$

We will denote the image of $K_p \otimes
t_0^{r_0} \t^\r$ in $\K$
by $t_0^{r_0} \t^\r K_p$, where 
$\r = (r_1, \ldots r_n)$ and $\t^\r = 
t_1^{r_1}\ldots t_n^{r_n}$.
Note that $\K$ has the defining relations
$$ r_0 t_0^{r_0} \t^\r K_0 + r_1 t_0^{r_0} \t^\r K_1
+ \ldots + r_n t_0^{r_0} \t^\r K_n = 0. $$

The toroidal algebra is the Lie
algebra 
$$\hat \g = \dot \g \otimes \C [t_0^\pm,
t_1^\pm,\ldots, t_n^\pm] \oplus \K$$
with the bracket
$$ [g_1 \otimes f_1(t_0 \ldots t_n) , 
g_2 \otimes f_2(t_0 \ldots t_n) ] = 
[g_1, g_2] \otimes (f_1 f_2) + (g_1 | g_2)
\sum\limits_{p=0}^n d_p(f_1)f_2 K_p  \eqno{(1.1)}$$
and
$$[\hat\g, \K] = 0. \eqno{(1.2)}$$

Finally, we shall add certain 
derivations to $\hat \g$.
Specifically, let $\cal D$ be the Lie algebra of 
derivations of $\C [t_0^\pm, t_1^\pm,\ldots, t_n^\pm]$:
$${\cal D} = \bigl\{ \sum\limits_{p=0}^n
f_p(t_0,\ldots,t_n)d_p \big|
f_0,\ldots,f_n \in \C [t_0^\pm, t_1^\pm,\ldots, 
t_n^\pm] \bigr\} .$$
 
It is a general fact, that a derivation acting on a 
Lie algebra can be lifted in a unique way to a 
derivation of the universal central extension of
this Lie algebra [BM].
Thus the natural action of $\cal D$
on $\tilde \g$
$$ f_1 (t_0,\ldots,t_n) d_p ( g \otimes
f_2(t_0,\ldots,t_n) = g \otimes f_1 d_p (f_2) 
\eqno{(1.3)}$$
has a unique extension to $\hat \g$. We shall denote
the lift of  $f (t_0,\ldots,t_n) d_p$ by
$f (t_0,\ldots,t_n) D_p$. Its action on the
subspace $\tilde \g$ is unchanged, while the action
on $\K$ is given by formula [EM]:
$$ f_1 D_a (f_2 K_b)  = f_1 D_a(f_2) K_b +
\delta_{ab} \sum\limits_{p=0}^n f_2 D_p(f_1) K_p . 
\eqno{(1.4)}$$

Consider a subalgebra $\D_+$ in $\D$:
$${\cal D}_+ = \bigl\{ \sum\limits_{p=1}^n
f_p(t_0,\ldots,t_n)d_p \big|
f_1,\ldots,f_n \in \C [t_0^\pm, t_1^\pm,\ldots, 
t_n^\pm] \bigr\} .$$
We will be working with the algebra $\g$ 
which is a deformation of the semidirect
product of $\hat \g$ with ${\cal D}_+$:
$$ \g = \dot \g \otimes \C [t_0^\pm, t_1^\pm,\ldots, 
t_n^\pm] \oplus \K \oplus {\cal D}_+ .$$
Here multiplication in $\hat\g$ is given by 
(1.1) and (1.2), the action of $\D_+$ on $\tg$
is given by (1.3) and on $\K$ by (1.4).
The Lie bracket in $\D_+$ is the following:
$$[t_0^{r_0} \t^\r D_a, t_0^{m_0} \t^\m D_b] = 
m_a t_0^{r_0+m_0} \t^{\r+\m} D_b -
r_b t_0^{r_0+m_0} \t^{\r+\m} D_a -
m_a r_b \bigl\{ \sum\limits_{p=0}^n 
r_p t_0^{r_0+m_0} \t^{\r+\m} K_p \bigr\} .$$
The last term in this formula is the correction term to the 
ordinary Lie bracket in $\D$.
This abelian deformation of the Lie algebra of
vector fields on a torus is a generalization
of the Virasoro algebra and was introduced in [EM]. 



\

\

{\bf 2. Principal realization of the basic 
representation of affine Lie algebra.}

\


When $n = 0$ the algebra $\g$ constructed above yields 
the derived affine Kac-Moody algebra.
We set $n = 0$ throughout this section.  
In this case $\K$ is one-dimensional and is spanned
by $K_0$, while $\D_+$ is trivial, so
$$\g = \dot\g \otimes \C[ t_0, t_0^{-1} ] 
\oplus \C K_0. $$

Let us recall the principal realization of the basic 
representation of $\g$ (see [Kac] for details). 
The starting point of this
construction is the principal Heisenberg subalgebra
in $\g$. 

Let $\{ \alpha_1, \ldots, \alpha_\ell \}$ be simple
roots and $\theta$ be the highest positive root in
$\dot \Delta$. We define the height
for a root $\beta = \sum\limits_{i=1}^\ell
k_i \alpha_i$ as
$$\hgt (\beta) = \sum\limits_{i=1}^\ell k_i.$$
Note that $\hgt(\theta) = h - 1$, where $h$ is a 
Coxeter number of $\dot\g$ [Kos].

Consider the principle gradation of $\tilde \g$ defined by
$$\deg (\dot\g^{\alpha_1} \otimes 1) = \ldots = 
\deg (\dot\g^{\alpha_\ell} \otimes 1) =
\deg (\dot\g^{-\theta} \otimes t_0) = 1 , $$
$$\deg (\dot\g^{-\alpha_1} \otimes 1) = \ldots = 
\deg (\dot\g^{-\alpha_\ell} \otimes 1) =
\deg (\dot\g^{\theta} \otimes t_0^{-1}) = -1 . $$
We choose nonzero root vectors
$e_0 \in \dot\g^{-\theta} \otimes t_0,
e_1 \in \dot\g^{\alpha_1} \otimes 1, \ldots,
e_\ell \in  \dot\g^{\alpha_\ell}$ and form
the element $\ol e = \sum\limits_{i=0}^\ell e_i
\in \tilde \g$. Since $\ol e$ is of degree $1$ then 
its centralizer $\tilde \s$ in $\tilde g$ is homogeneous 
with respect to the gradation. 

Next, consider a projection 
$$\pi : \dot \g \otimes \C[ t_0, t_0^{-1} ] 
\rightarrow \dot\g {\rm \ \ \ \ by \ \ \ \ } t_0 \mapsto 1
.$$
Note that
$\deg(\dot\g^{-\theta}\otimes t_0) = 1$  and
$\deg(\dot\g^{-\theta}\otimes 1) = ht(-\theta)
= -h + 1 $, while $\pi(\dot\g^{-\theta}\otimes t_0) =
\pi(\dot\g^{-\theta}\otimes 1)$,
and thus the principal $\Z$-gradation of $\tilde \g$
induces the $\Z_h$-gradation of $\dot\g$:
$$ \dg = \sum\limits_{j\in\Z_h} \dg_j .$$

The element $\pi(\ol e)$ is regular in $\dg$, hence
its centralizer $\dot\s = \pi(\tilde \s)$ in $\dg$ is
a Cartan subalgebra, which implies that both $\dot \s$ and 
$\tilde \s = \dot\s \otimes \C[t_0, t_0^{-1}]$
are abelian. 

The preimage $\s = \dot\s \otimes \C[t_0, t_0^{-1}] \oplus \C K_0$
of $\tilde\s$ in $\g$ is
an infinite-dimensional Heisenberg algebra which is 
called the principal Heisenberg subalgebra of $\g$.
This subalgebra is the cornerstone of the construction
of the vertex operator representation of $\g$.
The scheme of this construction is the following:
one starts with the standard irreducible representation 
of $\s$ and then magically one can lift this
representation to the whole of $\g$. Moreover
the action of $\g$ is prescribed and given by
vertex operators. Finally one has only to check in one
way or another that this construction works. 

The irreducible representation of $\g$ constructed 
in this way is of the highest weight  $\Lambda_0$,
where the linear functional $\Lambda_0$ is given 
(we consider the simply-laced case) by
$$ \Lambda_0 ( \nu^{-1} (\alpha_i ) \otimes t_0^0 ) = 0,
{\rm \ \ \ } \Lambda_0 ( K_0 ) = 1.$$
This highest weight representation is called the
basic representation of $\g$.

The Cartan subalgebra $\dot\s$ is homogeneous with
respect to the principle $\Z_h$-gradation:
$$ \dot\s = \sum\limits_{j\in\Z_h} \dot\s_j .$$

Since $(\dg_i | \dg_j ) = 0$ unless $i+j = 0$ (mod $h$)
and $(\cdot |\cdot )$ is nondegenerate on $\dot\s$ then
one can choose a basis in $\dot\s$: $\{T_1, T_2, 
\ldots, T_\ell \}$ such that $T_i \in \dg_{m_i}$,
where $1\leq m_1 \leq m_2 \leq \ldots \leq m_\ell <h$
and
$$ (T_i | T_{\ell + 1 -j}) = h \delta_{ij} .$$
The numbers $\{m_1,\ldots,m_\ell \}$ are called the
exponents of $\dg$ [Kos]. 
Note that $\dot\s \cap \dot\h = (0)$,
and therefore zero is not an exponent. Also, $m_{\ell + 1 - i} = 
h - m_i$. 

The principal Heisenberg subalgebra $\s$ 
is spanned by $T_i\otimes t_0^j$, $j\in\Z, 
i=1,\ldots,\ell $ and $K_0$. We can now write 
the multiplication in $\s$ explicitly:
$$[T_{i_1} \otimes t_0^{j_1}, T_{\ell + 1 -i_2}
\otimes t_0^{j_2}] = h j_1 \delta_{i_1 i_2}
\delta_{j_1, -j_2}.$$

 It turns out that it is more convenient to work with a slightly
different realization of $\g$ based on the
$\Z_h$-gradation of $\dg$. Consider
$$\g_s = \sum\limits_{j\in\Z} \dg_j \otimes s^j 
\oplus \C K_0 , $$
with the Lie bracket
$$[g_1\otimes s^i, g_2\otimes s^j] =
[g_1,g_2]\otimes s^{i+j} +
{i\over h} (g_1|g_2) \delta_{i,-j} K_0 ,$$
$$[K_0, \g_s] = 0 .$$

The proof of the following lemma is straightforward, so we
omit it.

{\bf Lemma 1}. 
{\it The Lie algebras $\g$ and $\g_s$ are
isomorphic and the isomorphism}
$$\psi : \dg\otimes \C [t_0,t_0^{-1}] \oplus \C K_0
\rightarrow \sum\limits_{j\in\Z} \dg_j \otimes s^j
 \oplus \C K_0 $$
{\it is given by }
$$\psi (e_\alpha \otimes t^i) = e_\alpha 
\otimes s^{\hgt(\alpha)+ih} ,$$
$$\psi (\nu^{-1}(\alpha) \otimes t^i) =
\nu^{-1}(\alpha) \otimes s^{ih} + \delta_{i,0}
{\hgt(\alpha) \over h} K_0 ,$$
$$\psi (K_0) = K_0 .$$

We shall identify $\g$ and $\g_s$ using this 
isomorphism.

In order to describe a basis in the principal Heisenberg subalgebra
$\psi(\s)$ we need to introduce the sequence $\{ b_i \}_{i\in\Z}$ such that
$b_{i+j\ell} = m_i + jh$ for $j\in\Z, i = 1,\ldots,\ell$.
Then $\psi (\s)$
is spanned by $\{ T_i \otimes s^{b_i} , K_0 \}, i\in \Z$.
Note that $b_{1-i} = - b_i$.

Let $\dot\Delta_s$ be the root system of $\dg$ with
respect to the Cartan subalgebra $\dot\s$.
For $\alpha\in\dot\Delta_s$ choose a root element
$$A^\alpha = \sum\limits_{j\in\Z_h} A^\alpha_j .$$
Since $\pi(\ol e)$ is a regular element of the Cartan
subalgebra then $[\pi(\ol e), A^\alpha] =
\lambda^\alpha A^\alpha$ with $\lambda^\alpha \neq 0$.
Hence $A^\alpha_j = \bigl( {{\rm ad} \pi (\ol e) \over
\lambda^\alpha} \bigr)^j A^\alpha_0$ and all the components
$A^\alpha_j$ are nonzero.
We let the indices $i,j$ in $T_i$ and $A^\alpha_j$ run
over $\Z$ by setting $T_{i_1} = T_{i_2}$ if $i_1 \equiv
i_2 (\mod \ell)$ and $A^\alpha_{j_1} = A^\alpha_{j_2}$
if $j_1 \equiv j_2 (\mod h)$.

Define constants $\lambda^\alpha_i = \alpha (T_i).$
Then 
$ [T_i , A^\alpha] = \lambda^\alpha_i A^\alpha$
 and
$ [T_i , A^\alpha_j] = \lambda^\alpha_i 
A^\alpha_{j + m_i} .$

Consider an automorphism $\sigma$ of $\dg$ such that
$$\sigma = \zeta^j \hbox{Id \ \ on \ } \dg_j ,$$
where $\zeta\in\C$ is the primitive h-root of 1.
For a root element $A^\alpha = \sum\limits_{j=1}^h A^\alpha_j$
we have
$$[T_i , \sigma(A^\alpha) ] =
 \sigma( [\sigma^{-1} (T_i), A^\alpha ]) =
 \sigma ([ \zeta^{-m_i} T_i , A^\alpha ]) =
 \zeta^{-m_i} \lambda^\alpha_i \sigma(A^\alpha) .$$
Thus $\sigma(A^\alpha) = \sum\limits_{j=1}^h \zeta^j A^\alpha_j $
is also a root element and $\sigma$ induces an automorphism
of $\dot\Delta_s$. Since all $A^\alpha_j$ are nonzero then
the length of each orbit of $\sigma$ in $\dot\Delta_s$ is $h$.

We may choose $\ell$ roots $\beta_1,\ldots,\beta_\ell
\in \dot\Delta_s$ such that $A^{\beta_1}_0,\ldots,A^{\beta_\ell}_0$
span $\dot\g_0$. Note that if two root elements belong to the same
orbit under the action of $\sigma$ then their zero componens
are proportional. Therefore every orbit of $\sigma$
contains exactly one of $\beta_1,\ldots,\beta_\ell$. 
Consequently, we may choose $\{ \sigma^j (A^{\beta_i}) \}, i=1,\ldots,\ell, 
j=1,\ldots,h$ as our family of the root elements.
The set $\{ A^{\beta_i}_j \otimes s^j , T_j \otimes s^{b_j} ,
K_0 \} , i=1,\ldots,\ell, j\in\Z$ forms a basis of $\g_s$.

{\it Remark.} The properties of the automorphism $\sigma$ were
studied by Kostant in the important paper [Kos]. He proved,
in particular, that $\sigma$ acts on the root system 
$\dot\Delta_s$ as a Coxeter transformation.
Remarkably, the complex coefficients $\lambda_i^\alpha$ could be
interpreted as the orthogonal projections of the roots on a real plane
invariant under the Coxeter transformation. These projections
were introduced by Coxeter in order to visualize the regular
polytopes of higher dimensions. In Chapters 12, 13 of [Co1] and
in section 12.5 of [Co2] the case of the root system of type $H_4$
is treated, which also gives the answer for $E_8$. 

Now we shall construct the standard representation
$(\phi, F)$ of $\s$. The space of this representation
(called the Fock space)
is the polynomial algebra in the infinitely many 
variables:
$$ F = \C[x_1, x_2, x_3, \ldots ].$$
For $i\geq 1$, $T_{1-i}\otimes s^{-b_i}$ is represented
by a multiplication operator:
$$\phi(T_{1-i}\otimes s^{-b_i}) = b_i x_i ,$$
$T_i \otimes s^{b_i}$ is represented
by a differentiation operator:
$$\phi(T_i\otimes s^{b_i}) = {\d\over\d x_i} $$
and $\phi(K_0) =$ Id.

Indeed, the only relation to be checked is
$$[\phi (T_i \otimes s^{b_i}) , 
\phi (T_{1-j} \otimes s^{b_{1-j}} )] = 
{1\over h} (T_i |T_{1-j}) b_i \delta_{b_ib_j} \phi(K_0) .$$
But as it can be easily seen, both expressions are equal to 
$b_i \delta_{ij}$ Id.

One can lift this representation of $\s$ to
the whole $\g$ using vertex operators. We consider
the space $\g[[z,z^{-1}]]$ of formal Laurent series
in a variable $z$ with coefficients in $\g$ and the space 
$\End(F)[[z,z^{-1}]]$
of series  with coefficients
in $\End(F)$ (see Chapter 2 in [FLM] for details).
The adjoint action of $\g$ on $\g[[z,z^{-1}]]$ is
well-defined.


 Every representation $\phi : \g \rightarrow \End(F)$
defines a homomorphism $\phi : \g[[z,z^{-1}]]
 \rightarrow \End(F)[[z,z^{-1}]]$.

The following theorem shows how to lift the standard
representation of the principal Heisenberg subalgebra
to $\g$:

\noindent 
{\bf Theorem 2} ([Kac]). 
{\it There exists a representation
$$ \phi : \g \rightarrow \End(\C[x_1, x_2, \ldots])$$
such that for} $i\in\N$
$$\phi(T_{1-i}\otimes s^{-b_i}) = b_i x_i ,$$
$$\phi(T_i\otimes s^{b_i}) = {\d\over\d x_i} , $$
$$\phi(K_0) = {\rm Id} ,$$
$$ \sum\limits_{j\in\Z} \phi(A^\alpha_j \otimes s^j ) z^{-j}
= A^\alpha (z) , {\it \ where \ }$$
$$ A^\alpha (z) = \Lambda_0 (A^\alpha_0 \otimes s^0)
exp(\sum_{i=1}^\infty \lambda^\alpha_i  z^{b_i} x_i )
exp( - \sum_{i=1}^\infty \lambda^\alpha_{1-i}
 {z^{-b_i}\over b_i} {\d\over\d x_i} ) . $$

 Since $A^\alpha_j $ together with
$T_i\otimes s^{b_i}$ and $K_0$ span $\g$ then
the above formulas completely determine this 
representation.


We will use extensively the following Laurent series:
$$ \delta(z) = \sum_{j\in\Z} z^j   \hbox{ \ \ and \ \ }
D\delta(z) = \sum_{j\in\Z}j z^j.$$
The first of these is the formal analogue of
the delta function.

For an element $B = \sum\limits_{j\in \Z_h} B_j
\in \dg$ denote by $B(z)$ the formal series
$\sum\limits_{j\in \Z} \phi(B_j \otimes s^j) z^{-j}
\in \End(F)[[z,z^{-1}]]$.

{\bf Corollary 3.}
 {\it Let $A = \sum\limits_{j=1}^h A_j ,
B = \sum\limits_{j=1}^h B_j \in \dg$ and let
$C^j = [A_j, B] \in \dg$. Then
$$ [A(z_1), B(z_2)] =
\sum_{j=1}^h \bigl( C^j (z_2) + {j \over h}(A_j |B_{-j})
\bigr)
 \zba^j \delta\bigl( \zba^h \bigr) +
\sum_{j=1}^h (A_j|B_{-j}) \zba^j D\delta \bigl( \zba^h \bigr).$$
}

{\it Proof of the Corollary.}
$$ [A(z_1), B(z_2)] =
\sum_{j\in\Z}  \sum_{i\in\Z}
[\phi(A_j\otimes s^j), \phi(B_i \otimes s^i)] z_1^{-j} z_2^{-i} =$$
$$\sum_{j\in\Z}  \sum_{i\in\Z}
\bigl\{ \phi([A_j,B_i]\otimes s^{j+i}) +
{1\over h} (A_j|B_i) j \delta_{j,-i} \phi(K_0) \bigr\}
 z_1^{-j} z_2^{-i} =$$
$$\sum_{j\in\Z}  \sum_{k=i+j\in\Z}
 \phi([A_j,B_{k-j}]\otimes s^k)  z_1^{-j} z_2^{j-k} +
{1\over h} \sum_{j\in\Z}(A_j|B_{-j}) j \zba^j =$$
$$ \sum_{j\in\Z} C^j(z_2) \zba^j +
{1\over h} \sum_{j_1=1}^h  \sum_{{j_2\in\Z}\atop {j = j_1 + hj_2}}
(A_{j_1}|B_{-j_1}) (j_1 + hj_2) \zba^{j_1 + hj_2} =$$
$$ \sum_{j_1=1}^h  \bigl( C^{j_1}(z_2)  +
{j_1\over h} (A_{j_1}|B_{-j_1})
\bigr) \zba^{j_1} \delta\bigl( \zba^h \bigr) +
\sum_{j_1=1}^h
(A_{j_1}|B_{-j_1}) \zba^{j_1} D\delta\bigl( \zba^h \bigr).$$

\

\

{\bf 3. Construction of the vertex operator representation
for the toroidal Lie algebra.}

\

In this section we construct a vertex operator representation
for the toroidal Lie algebra that generalizes the principal
realization of the basic representation of affine Lie algebra.
Again it will be convenient to replace $\g$ with
$$\g_s = \sum\limits_{j\in\Z} \dg_j \otimes s^j
\C[t_1^\pm,\ldots,t_n^\pm]
\oplus \K \oplus \D_+ , $$
with the Lie bracket
$$ [g_1 \otimes f_1(s,\t) , 
g_2 \otimes f_2(s,\t) ] = 
[g_1, g_2] \otimes (f_1 f_2) + (g_1 | g_2)
\bigl\{ {1\over h} (s {d\over ds} f_1)f_2 K_0 + 
\sum\limits_{p=1}^n d_p(f_1)f_2 K_p \bigr\} , $$
$$ [ f_1 (s,\t) D_p ,
 g \otimes f_2(s,\t) ] = g \otimes f_1 (d_p f_2)$$
while the multiplication in $\D_+$ and its
action on $\K$ are the same as in $\g$. Here and below
we make identifications 
$s^{hr_0} \t^\r K_a = t_0^{r_0} \t^\r K_a$ and
$s^{hr_0} \t^\r D_b = t_0^{r_0} \t^\r D_b$, 
$a = 0, 1, \ldots, n , b = 1, \ldots, n$.

The following lemma is an immediate generalization of Lemma 1:
 
{\bf Lemma 4.}
{\it The Lie algebras $\g$ and $\g_s$ are
isomorphic and the isomorphism is given by }
$$\psi (e_\alpha \otimes t_0^{r_0}\t^\r) =
e_\alpha \otimes s^{\hgt(\alpha) + hr_0}\t^\r ,$$
$$\psi (\nu^{-1}(\alpha) \otimes t_0^{r_0}\t^\r) =
\nu^{-1}(\alpha) \otimes s^{hr_0}\t^\r +
{\hgt(\alpha) \over h}  s^{hr_0}\t^\r K_0 ,$$
$$\psi = \hbox{ Id \ on \ } \K \oplus \D_+ .$$

We shall identify $\g$ and $\g_s$ using this 
isomorphism.

The subalgebra $\s$ 
with the basis $\{ T_i \otimes s^{b_i} , s^{ih} D_p,
s^{ih} K_p, K_0 \}$ , \ $i\in\Z$, \ $p = 1,\ldots, n$
is the principal (degenerate) Heisenberg subalgebra 
of $\g$. 

 Indeed, $K_0$ is its central element and the multiplication
in $\s$ is given by
$$ [T_i \otimes s^{b_i} , T_{1-j} \otimes s^{-b_j} ] =
b_i \delta_{ij} K_0 ,$$
$$ [s^{ih} D_a , s^{jh} K_b] = \delta_{ab} i s^{(i+j)h} K_0 
= i \delta_{ab} \delta_{i, -j} K_0 ,$$
$$  [s^{jh} D_p ,  T_i \otimes s^{b_i} ] = 0,
 [s^{ih} D_a , s^{jh} D_b] = 0,$$
$$  [T_i \otimes s^{b_i} , s^{jh} K_p] = 0, 
 [s^{ih} K_a , s^{jh} K_b] = 0,$$
where $i,j \in\Z , a,b,p = 1,\ldots,n$. 
This subalgebra is degenerate as a Heisenbeg algebra
since $[D_p , K_p] = 0$.

The Heisenberg algebra $\s$ can be represented
on the Fock space
$$ F = \C [q_p^\pm, x_i, u_{pi}, v_{pi}]^{p=1,\ldots,n}_{i\in\N}$$
by differentiation and multiplication operators:
$$ \phi(T_i \otimes s^{b_i} ) = {\d\over\d x_i} {\rm \ , \ \ }
 \phi(T_{1-i} \otimes s^{-b_i} ) = b_i x_i  ,$$
$$ \phi( s^{ih} D_p ) = {\d\over\d u_{pi}} {\rm \ , \ \ }
\phi( s^{-ih} D_p ) = i v_{pi} ,$$
$$ \phi( s^{ih} K_p ) = {\d\over\d v_{pi}} {\rm \ , \ \ }
\phi( s^{-ih} K_p ) = i u_{pi} ,$$
$$ \phi( D_p ) = q_p {\d\over\d q_{p}} {\rm \ , \ \ }
 \phi( K_p ) = 0 ,$$
$$ \phi (K_0) = {\rm Id} ,$$
where $i\geq 1$ and $p = 1,\ldots,n$.
Our goal is to extend this representation of $\s$ to $\g$.
Consider the following elements of $\g[[z,z^{-1}]]$:
$$ \sum_{j\in\Z} s^{jh} \t^\r K_0 z^{-jh} {\rm \ \ \ and}
\sum_{j\in\Z} A^\alpha_j \otimes s^j \t^\r z^{-j} .$$
Note that
$$ [ T_i \otimes s^{b_i} , 
\sum_{j\in\Z} s^{jh} \t^\r K_0 z^{-jh}] = 0,$$
$$ [ s^{ih} K_p , \sum_{j\in\Z} s^{jh} \t^\r K_0 z^{-jh}] = 0,$$
$$ [ s^{ih} D_p , \sum_{j\in\Z} s^{jh} \t^\r K_0 z^{-jh}] = 
\sum_{j\in\Z} r_p s^{(i+j)h} \t^\r K_0 z^{-jh} =$$
$$\sum_{k=i+j\in\Z} r_p s^{kh} \t^\r K_0 z^{-kh+ih} =
r_p z^{ih} \sum_{j\in\Z} s^{jh} \t^\r K_0 z^{-jh}.$$

The theory of vertex operators suggests (see Lemma 14.5 in [Kac])
that \break
$\sum\limits_{j\in\Z} s^{jh} \t^\r K_0 z^{-jh}$ should be
represented by
$$ K_0 (z,\r) = \q^\r 
\exp { \sum_{p=1}^n r_p \sum_{j\geq 1} z^{jh} u_{pj} }
\exp { - \sum_{p=1}^n r_p \sum_{j\geq 1} {z^{-jh}\over j} 
{\d\over\d v_{pj} }} .$$
Here $\q^\r = q_1^{r_1}\ldots q_n^{r_n}$.

In a similar way the commutator relations
{\baselineskip=9pt
$$ [ T_i \otimes s^{b_i} , 
\sum_{j\in\Z} A^\alpha_j \otimes s^j \t^\r z^{-j}] = 
\sum_{j\in\Z} \lambda^\alpha_i A^\alpha_{j+b_i}
\otimes s^{j+b_i} \t^\r z^{-j} = $$
$$ \lambda^\alpha_i \sum_{k=j+b_i\in\Z}  A^\alpha_k
\otimes s^k \t^\r z^{-k+b_i} =
\lambda^\alpha_i  z^{b_i}
\sum_{j\in\Z} A^\alpha_j \otimes s^j \t^\r z^{-j} ,$$
$$ [ s^{ih} K_p , 
\sum_{j\in\Z} A^\alpha_j \otimes s^j \t^\r z^{-j} ] = 0,$$
$$ [ s^{ih} D_p , 
\sum_{j\in\Z} A^\alpha_j \otimes s^j \t^\r z^{-j} ] = 
\sum_{j\in\Z} r_p A^\alpha_j \otimes s^{j+ih} \t^\r z^{-j} =$$
$$ r_p \sum_{k =j+ih\in\Z}  A^\alpha_j \otimes s^k \t^\r z^{-k+ih} =
r_p z^{ih} \sum_{j\in\Z} A^\alpha_j \otimes s^j \t^\r z^{-j}$$
suggest that 
$\sum\limits_{j\in\Z} A^\alpha_j \otimes s^j \t^\r z^{-j}$
should be represented by
$$A^\alpha (z,\r) = $$
$$ \q^\r \Lambda_0 (A^\alpha_0 \otimes s^0)
\exp{\sum_{i=1}^\infty \lambda^\alpha_i  z^{b_i} x_i }
\exp{ - \sum_{i=1}^\infty \lambda^\alpha_{1-i}
 {z^{-b_i}\over b_i} {\d\over\d x_i} } \times$$
$$ \times \exp { \sum_{p=1}^n r_p \sum_{j\in\Z} z^{jh} u_{pj} }
\exp { - \sum_{p=1}^n r_p \sum_{j\in\Z} {z^{-jh}\over j} 
{\d\over\d v_{pj} }} = $$
$$ A^\alpha (z) K_0 (z,\r).$$
The last formula hints that for $B\in\dg$ we may try
to represent $\sum\limits_{j\in\Z} B_j \otimes s^j \t^\r z^{-j}$
by
$$ B(z,\r) = B(z) K_0(z,\r).$$
}
Note that $B(z)$ is known from the affine case (Theorem 2).

In fact, we shall see that the same approach is valid
even for $\K$ and $\D_+$, but for these we should first
discuss the concept of the normal ordering.  

Let $X(z) = \sum\limits_{i\in\Z} X_i z^i ,
 Y(z) = \sum\limits_{j\in\Z} Y_j z^j \in \End(F)[[z,z^{-1}]]$.
We say that the product $X(z)Y(z)$ exists if for every $k\in\Z$
and for every $v\in F$ the sum
$$ \sum_{i\in\Z} X_i Y_{k-i} v$$
has finitely many nonzero terms. If this is the case then
$$X(z)Y(z) = \sum_{k\in\Z} \bigl( \sum_{i\in\Z} X_i Y_{k-i} 
\bigr) z^k .$$

It is possible that $X(z)Y(z)$ exists while $Y(z)X(z)$ does not.
For example, this happens for 
$$X(z) = \sum_{i\geq 1} x_i z^i {\rm \ \ and \ \ }
Y(z) = \sum_{i\geq 1} {\d\over\d x_i} z^{-i} .$$
To improve the situation we define the normal ordering
for the product of operators on $F$. For a pseudodifferential
operator $P\in\End(F)$ the normal ordering ${\bf :}x_i P{\bf :}$ is defined
as $x_i P$, while the normal ordering of the product
${\bf :}{\d\over\d x_i} P{\bf :}$ is defined as $P {\d\over\d x_i}$.
Note that for $X(z)$ and $Y(z)$ from the above example
${\bf :}X(z)Y(z){\bf :} = {\bf :}Y(z)X(z){\bf :} =  X(z)Y(z)$.

From the action of the principle Heisenberg subalgebra on the Fock
space we see that the series
 $\sum\limits_{j\in\Z} s^{jh} K_p z^{-jh}$
and $\sum\limits_{j\in\Z} s^{jh} D_p z^{-jh}$ are represented by
$$ K_p(z) = \sum_{i\geq 1} i u_{pi} z^{ih} +
 \sum_{i\geq 1} {\d\over\d v_{pi}} z^{-ih} $$
and                   
$$ D_p(z) = \sum_{i\geq 1} i v_{pi} z^{ih} +
q_p {\d\over\d q_p} + \sum_{i\geq 1} {\d\over\d u_{pi}} z^{-ih} .$$ 

Using the analogy with $A^\alpha (z,\r)$ we shall represent
$\sum\limits_{j\in\Z} s^{jh} \t^\r K_p z^{-jh}$  by
$$ K_p (z,\r) = K_p(z) K_0(z,\r)$$
and $\sum\limits_{j\in\Z} s^{jh} \t^\r D_p z^{-jh}$ by
$$ D_p (z,\r) = {\bf :} D_p(z) K_0 (z,\r) {\bf :} 
\hbox{\ \ }.$$
Note that we use the normal ordering in the case of $D_p(z,\r)$
in order for the product to exist.

We summarize this discussion in our main theorem:

{\bf Theorem 5.} 
{\it There exists a representation $\phi$ of the toroidal
Lie algebra 
$$\dg \otimes \C[s^\pm, t_1^\pm,\ldots,t_n^\pm] \oplus \K \oplus \D_+ $$
 on the Fock space
$$F = \C [q_p^\pm, x_i, u_{pi}, v_{pi}]^{p=1,\ldots,n}_{i\in\N} $$
such that for} $i=1,\ldots,\ell, p=1,\ldots,n ,
\alpha\in\dot\Delta_s, \r\in\Z^n$

$$\sum_{j\in\Z} \phi( s^{jh} \t^\r K_0) z^{-jh} = K_0 (z, \r) = 
\q^\r \exp { \sum_{p=1}^n r_p \sum_{j\geq 1} z^{jh} u_{pj} }
\exp { - \sum_{p=1}^n r_p \sum_{j\geq 1} {z^{-jh}\over j} 
{\d\over\d v_{pj} }} , \eqno{(3.1)}$$
$$\sum_{j\in\Z} \phi( T_i \otimes s^{m_i+jh} \t^\r) z^{-m_i-jh} =
T_i (z,\r) =$$
$$ \{ \sum_{j\geq 1} (jh - m_i) z^{jh - m_i} x_{j\ell+1-i} +
\sum_{j\geq 0} z^{-jh-m_i} {\d\over\d x_{j\ell + i}} \} K_0(z,\r) ,
 \eqno{(3.2)}$$
$$ \sum\limits_{j\in\Z} \phi(A^\alpha_j \otimes s^j \t^\r ) z^{-j}
= A^\alpha (z,\r)  = $$
$$\Lambda_0 (A^\alpha_0 \otimes s^0 )
exp(\sum_{i=1}^\infty \lambda^\alpha_i  z^{b_i} x_i )
exp( - \sum_{i=1}^\infty \lambda^\alpha_{1-i}
 {z^{-b_i}\over b_i} {\d\over\d x_i} ) K_0 (z,\r) , \eqno{(3.3)} $$
$$\sum_{j\in\Z} \phi( s^{jh} \t^\r K_p) z^{-jh} = K_p (z, \r) =
\{ \sum_{i\geq 1} i u_{pi} z^{ih} +
 \sum_{i\geq 1} {\d\over\d v_{pi}} z^{-ih} \}K_0(z,\r) ,  \eqno{(3.4)}$$ 
$$\sum_{j\in\Z} \phi( s^{jh} \t^\r D_p) z^{-jh} = D_p (z, \r) =
{\bf :}\{ \sum_{i\geq 1} i v_{pi} z^{ih} +
q_p {\d\over\d q_p} + \sum_{i\geq 1} {\d\over\d u_{pi}} z^{-ih} \}
K_0(z,\r){\bf :} \hbox{\ } . \eqno{(3.5)}$$

\

\

{\bf 4. Proof of the main theorem.}

\

First of all, let us check that formulas (3.1) - (3.5) define a linear
map $\phi: \g \rightarrow \End(F)$. The linear dependencies
between the momenta at the left hand side are of the form
$$ \phi( A^{\sigma^k(\alpha)}_j \otimes s^j \t^\r) = 
 \zeta^{kj} \phi( A^\alpha_j \otimes s^j \t^\r) $$
and 
$$ j \phi( s^{jh} \t^\r K_0 ) + \sum_{p=1}^n r_p \phi( s^{jh} \t^\r K_p ) = 0.$$
These dependencies extend to the following relations for the
corresponding series:
$$ \sum_{j\in\Z} \phi( A^{\sigma^k(\alpha)}_j \otimes s^j \t^\r) z^{-j} =
\sum_{j\in\Z} \phi( A^\alpha_j \otimes s^j \t^\r)(\zeta^{-k} z)^{-j}$$
and
$$ - {1\over h} \dz {} \sum_{j\in\Z} \phi( s^{jh} \t^\r K_0) z^{-jh} 
 + \sum_{p=1}^n r_p \sum_{j\in\Z} \phi( s^{jh} \t^\r K_p) z^{-jh} = 0,$$
where $\dz = z {\d\over\d z}$.
We have to show that the same relations hold for the right hand sides
of (3.1) - (3.5). Indeed, noting that $\lambda^{\sigma^k (\alpha)}_i =
\zeta^{-km_i} \lambda^\alpha_i$ we obtain
$$ A^{\sigma^k (\alpha)} (z, \r) = \Lambda_0 (A^{\sigma^k(\alpha)}_0)
exp(\sum_{i=1}^\infty \lambda^\alpha_i \zeta^{-km_i} z^{b_i} x_i )
exp( - \sum_{i=1}^\infty \lambda^\alpha_{1-i} \zeta^{km_i}
 {z^{-b_i}\over b_i} {\d\over\d x_i} ) K_0 (z,\r) =$$ 
$$ \Lambda_0 (A^\alpha_0)
exp(\sum_{i=1}^\infty \lambda^\alpha_i (\zeta^{-k} z)^{b_i} x_i )
exp( - \sum_{i=1}^\infty \lambda^\alpha_{1-i} {(\zeta^{-k}
 z)^{-b_i}\over b_i} {\d\over\d x_i} ) K_0 (\zeta^{-k} z,\r) =
A^\alpha(\zeta^{-k}z, \r) .$$
The verification of the second relation is also straightforward:
$$ \dz K_0 (z, \r) = 
\dz \q^\r \exp { \sum_{p=1}^n r_p \sum_{j\geq 1} z^{jh} u_{pj} }
\exp { - \sum_{p=1}^n r_p \sum_{j\geq 1} {z^{-jh}\over j} 
{\d\over\d v_{pj} }} =$$
$$ \sum_{p=1}^n r_p \bigl\{ \sum_{j\geq 1} z^{jh} jh u_{pj} +
\sum_{j\geq 1} z^{-jh} h {\d\over\d v_{pj}} \bigr\} K_0 (z, \r) = $$
$$ h \sum_{p=1}^n r_p K_p (z, \r) .$$

Observing that the momenta of the series in the left hand sides of
(3.1) - (3.5) span $\g$ we conclude that the linear map $\phi :
\g \rightarrow \End (F)$ is well-defined. We need to show now that
$\phi$ is a homomorphism of Lie algebras.

The following Lemma and its Corollary which we state without proof
are very useful for the computations with formal series.

{\bf Lemma 6} (cf. Proposition 2.2.2 in [FLM]).
{\it Let} 
$$\delta(z) = \sum_{k\in\Z} z^k  \hbox{\rm \ \ \ \ and \ \ \ }
D\delta(z) = D_z \delta (z) = \sum_{k\in\Z} k z^k .$$
{\it If the products in the left hand sides exist then the following
equalities hold:}
$$ X(z_1) \delta (\zba) = X(z_2) \delta (\zba) , \leqno{(i)}$$
$$ X(z_1) D\delta (\zba) = X(z_2) D\delta (\zba) +
D_{z_2} ( X(z_2) ) \delta(\zba) , \leqno{(ii)}$$


{\bf Corollary 7. }
{\it Let $X(z) = Y(z^h)$. The following equalities hold
provided the products in the left hand sides exist:}
$$ X(z_1) \delta (\zba^h) = X(z_2) \delta (\zba^h) , \leqno{(i)}$$
$$ X(z_1) D\delta (\zba^h) = X(z_2) D\delta (\zba^h) +
{1\over h} D_{z_2} ( X(z_2) ) \delta(\zba^h) , \leqno{(ii)}$$
  
\
 
Note that $A^\alpha(z,\r) = A^\alpha(z) K_0(z,\r)$ and
$T_i (z, \r) = T_i (z) K_0(z,\r)$ , hence for every $B\in\dg$ 
we have $B(z,\r) = B(z) K_0 (z,\r)$.

In order \ to verify the commutator relations \ between the elements of
\break 
$\sum\limits_{j\in\Z}\dg_j \otimes s^j \C[t_1^\pm,\ldots,t_n^\pm]$ 
we have to show that for every $A, B \in \dg$
$$\sum_{j,i\in\Z} \phi([A_j \otimes s^j \t^\r, 
B_i \otimes s^i \t^\m]) z_1^{-j} z_2^{-i} = 
\sum_{j,i\in\Z} [\phi(A_j \otimes s^j \t^\r), \phi(B_i \otimes s^i \t^\m)]
z_1^{-j} z_2^{-i} . \eqno{(4.1)}$$
Indeed, let $C^j = [A_j , B]$. Then 
$$\sum_{j,i\in\Z} \phi([A_j \otimes s^j \t^\r, B_i \otimes s^i \t^\m]) 
z_1^{-j} z_2^{-i} = 
\hbox{\hskip 7cm} $$
$$\eqalign{ \hbox{\hskip 0.5cm}
&= \sum_{j,i\in\Z} \phi([A_j, B_i] \otimes s^{j+i} \t^{\r+\m}) z_1^{-j} z_2^{-i} +
\cr
&\quad
\sum_{j,i\in\Z} (A_j | B_i) \phi(s^{j+i} \t^{\r+\m} ({j\over h}K_0 +
\sum_{p=1}^n r_p K_p )) =
\cr   
&= \sum_{j\in\Z} \sum_{k=j+i\in\Z} \phi([A_j, B_{k-j}] \otimes s^k \t^{\r+\m})
z_1^{-j} z_2^{j-k} +
\cr 
&\quad
\sum_{j_1=1}^h \sum_{{j_2\in\Z}\atop{j=j_1+hj_2}}
\sum_{{i_2\in\Z}\atop{i=-j_1+hi_2}} (A_{j_1}|B_{-j_1}) 
{j_1 + hj_2\over h}
\phi(s^{h(j_2+i_2)} \t^{\r+\m} K_0 )
 z_1^{-j_1-hj_2} z_2^{j_1-hi_2} + 
\cr
&\quad
\sum_{j_1=1}^h \sum_{{j_2\in\Z}\atop{j=j_1+hj_2}}
\sum_{{i_2\in\Z}\atop{i=-j_1+hi_2}} (A_{j_1}|B_{-j_1}) 
\sum_{p=1}^n r_p \phi(s^{h(j_2+i_2)} \t^{\r+\m} K_p )
 z_1^{-j_1-hj_2} z_2^{j_1-hi_2} = 
\cr 
&= \sum_{j_1=1}^h \sum_{j_2\in\Z} \sum_{k\in\Z} \phi( C^{j_1}_k \otimes
s^k \t^{\r+\m}) z_2^{-k} \zba^{j_1} \zba^{hj_2} +
\cr
&\quad
\sum_{j_1=1}^h \sum_{j_2\in\Z}
\sum_{k=j_2+i_2\in\Z}  (A_{j_1}|B_{-j_1}) {j_1 + hj_2\over h}
\phi(s^{hk} \t^{\r+\m} K_0 )
 z_1^{-j_1-hj_2} z_2^{j_1+hj_2-hk} + 
\cr
&\quad
\sum_{j_1=1}^h \sum_{j_2\in\Z}
\sum_{k=j_2+i_2\in\Z}  (A_{j_1}|B_{-j_1})  \sum_{p=1}^n r_p
\phi(s^{hk} \t^{\r+\m} K_p )
 z_1^{-j_1-hj_2} z_2^{j_1+hj_2-hk} = 
\cr 
&= \sum_{j_1=1}^h C^{j_1} (z_2, \r+\m) \zba^{j_1} \delta(\zba^h) +
\cr                               
&\quad
{1\over h} \sum_{j_1=1}^h j_1 (A_{j_1}|B_{-j_1}) K_0(z_2,\r+\m)
\zba^{j_1} \delta (\zba^h) +
\cr 
&\quad
\sum_{p=1}^n r_p \sum_{j_1=1}^h (A_{j_1}|B_{-j_1}) K_p(z_2,\r+\m)
\zba^{j_1} \delta (\zba^h) +
\cr
&\quad
\sum_{j_1=1}^h (A_{j_1}|B_{-j_1}) K_0(z_2,\r+\m)
\zba^{j_1} D\delta (\zba^h).
}$$

For the right hand side of (4.1) we use Corollary 3, Corollary 7
and the fact that $K_0(z,\r)K_0(z,\m) = K_0(z,\r+\m)$:
$$ \sum_{j,i\in\Z} [\phi(A_j \otimes s^j \t^\r), \phi(B_i \otimes s^i \t^\m)]
z_1^{-j} z_2^{-i} = [A(z_1,\r),B(z_2,\m)] = $$
$$ = [A(z_1)K_0(z_1,\r),B(z_2)K_0(z_2,\m)] = 
[A(z_1),B(z_2)] K_0(z_1,\r) K_0(z_2,\m) = $$
$$ \eqalign{
&= \sum_{j=1}^h C^j (z_2) \zba^j \delta(\zba^h) K_0(z_1,\r) K_0(z_2,\m) +
\cr
&\quad
{1\over h} \sum_{j=1}^h j (A_j|B_{-j}) \zba^j \delta(\zba^h) 
K_0(z_1,\r) K_0(z_2,\m) +
\cr                     
&\quad
\sum_{j=1}^h (A_j|B_{-j}) \zba^j D\delta(\zba^h) K_0(z_1,\r) K_0(z_2,\m) =
\cr 
&= \sum_{j=1}^h C^j (z_2) \zba^j \delta(\zba^h) K_0(z_2,\r) K_0(z_2,\m) +
\cr
&\quad
{1\over h} \sum_{j=1}^h j (A_j|B_{-j}) \zba^j \delta(\zba^h) 
K_0(z_2,\r) K_0(z_2,\m) +
\cr                     
&\quad
\sum_{j=1}^h (A_j|B_{-j}) \zba^j D\delta(\zba^h) K_0(z_2,\r) K_0(z_2,\m) +
\cr
&\quad
\sum_{j=1}^h (A_j|B_{-j}) \zba^j \delta(\zba^h) {1\over h} 
D_{z_2}( K_0(z_2,\r) ) K_0(z_2,\m) = 
\cr &= \sum_{j=1}^h C^j (z_2) \zba^j \delta(\zba^h) K_0(z_2,\r+\m) +
\cr
&\quad
{1\over h} \sum_{j=1}^h j (A_j|B_{-j}) \zba^j \delta(\zba^h) K_0(z_2,\r+\m) +
\cr
&\quad
\sum_{j=1}^h (A_j|B_{-j}) \zba^j D\delta(\zba^h) K_0(z_2,\r + \m) +
\cr
&\quad
\sum_{j=1}^h (A_j|B_{-j}) \zba^j \delta(\zba^h) 
\sum_{p=1}^n r_p K_p(z_2) K_0(z_2,\r) K_0(z_2,\m) = 
\cr}$$
$$\eqalign{
&= \sum_{j=1}^h C^j (z_2) \zba^j \delta(\zba^h) K_0(z_2,\r+\m) +
\cr
&\quad
{1\over h} \sum_{j=1}^h j (A_j|B_{-j}) \zba^j \delta(\zba^h) K_0(z_2,\r+\m) +
\cr
&\quad
\sum_{p=1}^n r_p \sum_{j=1}^h (A_j|B_{-j}) \zba^j \delta(\zba^h) K_p(z_2,\r+\m) +
\cr 
&\quad
\sum_{j=1}^h (A_j|B_{-j}) \zba^j D\delta(\zba^h) K_0(z_2,\r + \m).
\cr }$$ 
Thus (4.1) holds. 

It is easy to see that operators $\phi(s^{jh} t^\r K_p), 
p=0,1,\ldots,n$ 
commute with 
\break
$\phi\bigl(\sum\limits_{j\in\Z}\dg_j\otimes 
s^j\C[t_1^\pm,\ldots,t_n^\pm]\oplus \K\bigr)$. 
This follows from the obvious equalities:
$$ [K_p(z_1,\r), A^\alpha (z_2,\m)] = [K_p(z_1,\r), T_i (z_2,\m)] = 0,$$
$$ [K_a(z_1,\r), K_b(z_2,\m)] = 0, {\rm \ \ \ } p, a, b = 0,1,\ldots,n.$$

In order to check the action of $\D_+$ on $\sum\limits_{j\in\Z} \dg_j
\otimes s^j \C[t_1^\pm,\ldots,t_n^\pm]$ it is sufficient to verify
the following equality for the generating series:
$$\sum_{j,i\in\Z} \phi( [s^{jh} \t^\r D_p , s^i \t^\m A^\alpha_i ] )
z_1^{-jh} z_2^{-i} = 
\sum_{j,i\in\Z} [ \phi( s^{jh} \t^\r D_p), \phi(A^\alpha_i \otimes s^i \t^\m )]
z_1^{-jh} z_2^{-i}. \eqno{(4.2)}$$
The corresponding equalities for $T_i \otimes s^{b_i} \t^\m$ and
$s^{ih} \t^\m K_p$ will follow since $s^{jh} \t^\r D_p$
acts as a derivation and $s^i \t^\m A^\alpha_i$ generate 
$\sum\limits_{j\in\Z} \dg_j \otimes s^j \C[t_1^\pm,\ldots,t_n^\pm]
\oplus \K$.

Let us prove that (4.2) holds.
$$\eqalign{
&\sum_{j,i\in\Z} \phi( [s^{jh} \t^\r D_p , s^i \t^\m A^\alpha_i ] )
z_1^{-jh} z_2^{-i} = \cr
&\sum_{j,i\in\Z} m_p \phi( s^{i+jh} \t^{\r+\m} A^\alpha_i )
z_1^{-jh} z_2^{-i} =\cr 
&\sum_{j\in\Z} \sum_{k=i+jh\in\Z} m_p \phi( s^k \t^{\r+\m} A^\alpha_k )
z_1^{-jh} z_2^{-k+jh} =\cr
& m_p A^\alpha(z_2,\r+\m) \delta(\zba^h) . \cr}$$

Next, we will show that the right hand side of (4.2) reduces to the same 
expression.
In the following computations it will be convenient to denote
$\phi( s^{jh} D_p ) = {\d\over\d u_{pj}}$ by $\gamma_p(j)$ ,
$\phi( s^{-jh} D_p ) = j v_{pj}$ by $\gamma_p(-j)$ and 
$ \phi( D_p ) = q_p {\d\over\d q_{p}}$ by $\gamma_p(0)$. Note that
$ [\gamma_p(j) , K_0(z,\m) ] = m_p z^{jh} K_0(z,\m)$ for $j\in\Z$.

$$\sum_{j,i\in\Z} [ \phi( s^{jh} \t^\r D_p),\phi(A^\alpha_i \otimes s^i \t^\m )]
z_1^{-jh} z_2^{-i} = [D_p(z_1,\r), A^\alpha(z_2,\m)] = $$
$$ [ {\bf :}\{ \sum_{j\in\Z} \gamma_p(j) z_1^{-jh} \} K_0(z_1,\r){\bf :},
A^\alpha(z_2) K_0(z_2,\m)] = $$
$$A^\alpha(z_2) [ {\bf :}\{ \sum_{j\in\Z} \gamma_p(j) z_1^{-jh} \} K_0(z_1,\r) {\bf :},
K_0(z_2,\m)] = $$
$$A^\alpha(z_2) 
[ \{ \sum_{j<0} \gamma_p(j) z_1^{-jh} \} K_0(z_1,\r), K_0(z_2,\m)] +$$
$$A^\alpha(z_2)  
[ K_0(z_1,\r)  \{ \sum_{j\geq 0} \gamma_p(j) z_1^{-jh} \} , K_0(z_2,\m) ] = $$
$$A^\alpha(z_2) [ \sum_{j<0} \gamma_p(j) z_1^{-jh}, K_0(z_2,\m)] K_0(z_1,\r) +
$$
$$A^\alpha(z_2) 
K_0(z_1,\r) [ \sum_{j\geq 0} \gamma_p(j) z_1^{-jh} , K_0(z_2,\m) ] = $$
$$A^\alpha(z_2) \sum_{j<0} z_1^{-jh} z_2^{jh} m_p K_0(z_2,\m) K_0(z_1,\r) +$$
$$A^\alpha(z_2) K_0(z_1,\r)\sum_{j\geq 0} z_1^{-jh} z_2^{jh} m_p K_0(z_2,\m)  = $$
$$ m_p A^\alpha(z_2) K_0(z_1,\r) K_0(z_2,\m) \delta( \zba^h ) =$$
$$ m_p A^\alpha(z_2) K_0(z_2,\r) K_0(z_2,\m) \delta( \zba^h ) =$$
$$ m_p A^\alpha(z_2,\r+\m) \delta( \zba^h ) .$$

From the above computation we can see that
$$[D_p(z_1,\r), K_0(z_2,\m)] = m_p K_0(z_2,\r+\m) \delta( \zba^h ) .$$

To complete  the proof of Theorem 5  we need to compute  the commutators
\break  
$[\phi(s^{jh} \t^\r D_a), \phi(s^{ih} \t^\m D_b)]$.  
We will check the following equality for the generating series:
$$ \sum_{j,i\in\Z} \phi( [s^{jh} \t^\r D_a, s^{ih} \t^\m D_b] ) 
z_1^{-jh} z_2^{-ih} =
\sum_{j,i\in\Z} [\phi(s^{jh} \t^\r D_a), \phi(s^{ih} \t^\m D_b)]
z_1^{-jh} z_2^{-ih} . \eqno{(4.3)}$$

To compute the left hand side we use multiplication in $\g$:
$$ \sum_{j,i\in\Z} \phi( [s^{jh} \t^\r D_a, s^{ih} \t^\m D_b] ) 
z_1^{-jh} z_2^{-ih} = \hbox{\hskip 6cm}$$
$$\eqalign{
&= \sum_{j,i\in\Z} \phi( m_a s^{(j+i)h} \t^{\r+\m} D_b -
r_b s^{(j+i)h} \t^{\r+\m} D_a ) z_1^{-jh} z_2^{-ih} -
\cr
&\quad
m_a r_b \phi ( s^{(j+i)h} \t^{\r+\m} 
\{ jK_0 + \sum_{p=1}^n r_p K_p \} ) z_1^{-jh} z_2^{-ih} =
\cr 
&= \sum_{j\in\Z} \sum_{k=j+i\in\Z} \big(
m_a \phi( s^{kh} \t^{\r+\m} D_b ) - r_b \phi( s^{kh} \t^{\r+\m} D_a) \bigl)
z_1^{-jh} z_2^{-kh+jh} -
\cr
&\quad
\sum_{j\in\Z} \sum_{k=j+i\in\Z}
m_a r_b \phi( s^{kh} \t^{\r+\m} \{ jK_0 + \sum_{p=1}^n r_p K_p \} ) \big)
z_1^{-jh} z_2^{-kh+jh} =
\cr
&= \bigl( m_a D_b(z_2,\r+\m) - r_b D_a(z_2,\r+\m) \bigr) \delta( \zba^h ) -
\cr
&\quad
m_a r_b \sum_{p=1}^n r_p K_p(z_2, \r+\m) \delta( \zba^h ) -
m_a r_b K_0(z_2,\r+\m) D \delta( \zba^h ) .
\cr}$$

 Observe that
$$ [D_a (z, \r), \gamma_b(i) ] = [ {\bf :}\{ \sum_{j\in\Z} \gamma_a(j) z^{-jh} \}
K_0 (z, \r){\bf :}, \gamma_b(i) ] = $$
$$ [ \sum_{j<0} \gamma_a(j) z^{-jh} K_0 (z, \r) , \gamma_b(i) ] +
[K_0 (z, \r) \sum_{j\geq 0} \gamma_a(j) z^{-jh} , \gamma_b(i) ] =$$
$$ \sum_{j<0} \gamma_a(j) z^{-jh} [ K_0 (z, \r) , \gamma_b(i) ] +
[K_0 (z, \r) , \gamma_b(i)] \sum_{j\geq 0} \gamma_a(j) z^{-jh} =$$
$$ - r_b z^{ih} \sum_{j<0} \gamma_a(j) z^{-jh} K_0 (z, \r) -
r_b z^{ih} K_0 (z, \r) \sum_{j\geq 0} \gamma_a(j) z^{-jh} =$$  
$$  - r_b z^{ih} D_a (z, \r) .$$

We use the last equality to compute the right hand side of (4.3):
$$\sum_{j,i\in\Z} [\phi(s^{jh} \t^\r D_a), \phi(s^{ih} \t^\m D_b)]
z_1^{-jh} z_2^{-ih} = [ D_a (z_1,\r), D_b(z_2,\m) ] = \hbox{\hskip 2cm}$$
$$= [ D_a (z_1,\r), \sum_{i<0} \gamma_b(i) z_2^{-ih} K_0 (z_2, \m) ] + 
[ D_a (z_1,\r), K_0 (z_2, \m) \sum_{i\geq 0} \gamma_b(i) z_2^{-ih} ] =
$$
$$\eqalign{ 
&= \sum_{i<0} [ D_a (z_1,\r), \gamma_b(i)] z_2^{-ih} K_0 (z_2, \m) + 
\sum_{i<0} \gamma_b(i) z_2^{-ih} [ D_a (z_1,\r), K_0 (z_2, \m) ] +
\cr 
&\quad
[ D_a (z_1,\r), K_0 (z_2, \m)] \sum_{i\geq 0} \gamma_b(i) z_2^{-ih} + 
 K_0 (z_2, \m) \sum_{i\geq 0} [ D_a (z_1,\r), \gamma_b(i)] z_2^{-ih}  =
\cr 
&= -r_b \sum_{i<0} D_a (z_1,\r) z_1^{ih} z_2^{-ih} K_0 (z_2, \m) +
\sum_{i<0} \gamma_b(i) z_2^{-ih} m_a K_0 (z_2, \r+\m) \delta(\zba^h) +
\cr
&\quad
m_a K_0 (z_2, \r+\m) \delta(\zba^h) \sum_{i\geq 0} \gamma_b(i) z_2^{-ih} -
r_b  K_0 (z_2, \m) \sum_{i\geq 0} D_a (z_1,\r) z_1^{ih} z_2^{-ih}  =
\cr
&= m_a D_b (z_2, \r+\m) \delta( \zba^h ) -
\cr
&\quad
r_b \sum_{i<0} \zab^{ih} \sum_{j<0} \gamma_a(j) z_1^{-jh} K_0(z_1,\r) 
K_0(z_2,\m)-
\cr
&\quad
r_b \sum_{i<0} \zab^{ih} K_0(z_1,\r) \sum_{j\geq 0} \gamma_a(j) z_1^{-jh}  
K_0(z_2,\m) -
\cr
&\quad
r_b  K_0 (z_2, \m) \sum_{i\geq 0} \sum_{j<0} \gamma_a(j) z_1^{-jh} K_0(z_1,\r)
 \zab^{ih} -
\cr 
&\quad
r_b  K_0 (z_2, \m) \sum_{i\geq 0} K_0(z_1,\r)\zab^{ih}
\sum_{j\geq 0} \gamma_a(j) z_1^{-jh} =
\cr
&= m_a D_b (z_2, \r+\m) \delta( \zba^h ) -
\cr
&\quad
r_b \sum_{i<0} \sum_{j<0} \zab^{ih} z_1^{-jh} \gamma_a(j) K_0(z_1,\r) 
K_0(z_2,\m)-
\cr
&\quad
r_b \sum_{i<0} \sum_{j\geq 0} \zab^{ih} z_1^{-jh} K_0(z_1,\r) [\gamma_a(j), 
K_0(z_2,\m)] -
\cr 
&\quad
r_b \sum_{i<0} \sum_{j\geq 0} \zab^{ih} z_1^{-jh} K_0(z_1,\r)  
K_0(z_2,\m) \gamma_a(j) -
\cr 
&\quad
r_b \sum_{i\geq 0} \sum_{j<0} z_1^{-jh} \zab^{ih} [K_0 (z_2, \m), \gamma_a(j)] K_0(z_1,\r)
  -
\cr 
&\quad
r_b  \sum_{i\geq 0} \sum_{j<0} z_1^{-jh} \zab^{ih} 
\gamma_a(j) K_0 (z_2, \m) K_0(z_1,\r) -
\cr 
&\quad
r_b  \sum_{i\geq 0} \sum_{j\geq 0} \zab^{ih} z_1^{-jh} K_0 (z_2, \m) K_0(z_1,\r)
\gamma_a(j) =
\cr }$$
$$\eqalign{
&= m_a D_b (z_2, \r+\m) \delta( \zba^h ) -
\cr
&\quad
r_b \delta( \zba^h) {\bf :}\{ \sum_{j\in\Z} \gamma_a(j)  z_2^{-jh} \} K_0(z_2,\r) K_0(z_2,\m){\bf :} -
\cr
&\quad
m_a r_b \sum_{i>0} \sum_{j\geq 0} \zba^{(i+j)h} K_0(z_1,\r) 
K_0(z_2,\m) +
\cr
&\quad
m_a r_b \sum_{i\leq 0} \sum_{j<0} \zba^{(i+j)h} K_0(z_1,\r) 
K_0(z_2,\m) =
\cr 
&= m_a D_b (z_2, \r+\m) \delta( \zba^h ) -
r_b D_a (z_2,\r+\m) \delta( \zba^h) -
\cr
&\quad
m_a r_b \sum_{k=i+j>0} \sum_{j=0}^{k-1} \zba^{kh} K_0(z_1,\r) K_0(z_2,\m) +
\cr
&\quad
m_a r_b \sum_{k=i+j<0} \sum_{j=k}^{-1} \zba^{kh} K_0(z_1,\r) K_0(z_2,\m) =
\cr
&= \bigl( m_a D_b (z_2, \r+\m) - r_b D_a (z_2,\r+\m) \bigr) \delta( \zba^h ) -
\cr
&\quad
m_a r_b \sum_{k\in\Z} k \zba^{kh} K_0(z_1,\r) K_0(z_2,\m) = 
\cr
&= \bigl( m_a D_b (z_2, \r+\m) - r_b D_a (z_2,\r+\m) \bigr) \delta( \zba^h ) -
\cr
&\quad
m_a r_b K_0(z_1,\r) K_0(z_2,\m) D\delta( \zba^h ) =
\cr
&= \bigl( m_a D_b (z_2, \r+\m) - r_b D_a (z_2,\r+\m) \bigr) \delta( \zba^h ) -
\cr
&\quad
{1\over h} m_a r_b D_{z_2} ( K_0 (z_2, \r) ) K_0(z_2,\m) \delta( \zba^h ) -
\cr
&\quad
m_a r_b K_0(z_2,\r) K_0(z_2,\m) D\delta( \zba^h )
 =
\cr
&= \bigl( m_a D_b (z_2, \r+\m) - r_b D_a (z_2,\r+\m) 
\bigr) \delta( \zba^h ) -
\cr
&\quad
m_a r_b \sum_{p=1}^n r_p K_p(z_2,\r+\m) \delta( \zba^h ) -
\cr
&\quad
m_a r_b K_0(z_2,\r+\m) D\delta( \zba^h ) .
\cr}$$

This establishes (4.3) and completes the proof of Theorem 5.

\

\vfill\eject

{\bf 5. Sugawara operators.}

\

In the representation we constructed the operators from $\D_+$ act
as derivations on the algebra $\hat\g = \dg\otimes 
\C[t_0^\pm, t_1^\pm,\ldots, t_n^\pm] \oplus \K$. 
It is possible to extend this action on $\D$, that is to
represent $ t_0^j \t^\r D_0$ by operators on $F$. For the affine case 
these are Sugawara operators.

We will work with the toroidal algebra realized
as $\sum\limits_{j\in\Z} \dg_j \otimes s^j \C[t_1^\pm,\ldots, t_n^\pm]  
\oplus \K$. For this realization
$$\psi (t_0^j \t^\r D_0) = {1\over h} s^{jh} \t^\r D_s -
{1\over h} \nu^{-1} (\rho) \otimes s^{jh} \t^\r ,$$
where $\rho\in\dot\h$ with $(\rho | \alpha_i) = 1$ for every simple
root $\alpha_i \in \dot\Delta$ and $D_s = s{d\over ds}$.

Since for $\alpha\in\Delta$ the elements
 $A^\alpha \otimes t_0^i \t^\m$ generate 
$\hat\g$ then we have to check that
$$ \psi( [t_0^j \t^\r D_0, A^\alpha \otimes t_0^i \t^\m] ) =
[ \psi( t_0^j \t^\r D_0), \psi(A^\alpha \otimes t_0^i \t^\m) ] .$$
Note that $A^\alpha\in\dg_{\hgt(\alpha)}$ and $(\rho|\alpha) = \hgt(\alpha)$.

We have
$$ \psi( [t_0^j \t^\r D_0, A^\alpha \otimes t_0^i \t^\m] ) =                 
i \psi( A^\alpha \otimes t_0^{i+j} \t^{\r+\m} ) = 
i A^\alpha \otimes s^{\hgt(\alpha)+h(i+j)} \t^{\r+\m} , $$
while
$$[ \psi( t_0^j \t^\r D_0), \psi(A^\alpha \otimes t_0^i \t^\m) ] =$$
$$[ {1\over h} s^{jh} \t^\r D_s -
{1\over h} \nu^{-1} (\rho) \otimes s^{jh} \t^\r,
A^\alpha \otimes s^{\hgt(\alpha)+ih} \t^\m ] =$$
$$ {1\over h} (\hgt(\alpha) + ih - (\rho|\alpha) )
A^\alpha \otimes s^{\hgt(\alpha)+h(i+j)} \t^{\r+\m} = $$
$$i A^\alpha \otimes s^{\hgt(\alpha)+h(i+j)} \t^{\r+\m} . $$
   
Thus it is sufficient to construct operators on $F$ corresponding to
$ s^{jh} \t^\r D_s $.

It will be convenient to denote in this section
$\phi(T_i\otimes s^{b_i})$ by $\tau(i)$ , $\phi(s^{jh}D_p)$ by $\gamma_p(j)$
and $\phi(s^{jh}K_p)$ by $\kappa_p(j)$.

In these notations
$$K_0(z,\r) = \q^\r 
\exp {\sum_{p=1}^n r_p \sum_{j\geq 1} {\kappa_p (-j)\over j} z^{jh} }
\exp { - \sum_{p=1}^n r_p \sum_{j\geq 1} {\kappa_p (j)\over j} z^{-jh} }
{\rm \ \ \ and}$$
$$A^\alpha(z) = \Lambda_0(A^\alpha_0 \otimes s^0)
\exp{ \sum_{i\geq 1} \lambda^\alpha_i {\tau(1-i)\over b_i} z^{b_i} }
\exp{ - \sum_{i\geq 1} \lambda^\alpha_{1-i} {\tau(i)\over b_i} z^{-b_i} } .$$

Nontrivial commutators in the principal Heisenberg subalgebra are
$$ [\tau(i), \tau(1-i)] = b_i  {\rm \ \ and \ \ }
[\gamma_p (j), \kappa_p (-j)] = j .$$

The action of the Heisenberg subalgebra on the vertex operators is determined by
$$ [\tau(i), A^\alpha(z)] = \lambda^\alpha_i z^{b_i} A^\alpha(z) ,$$
$$ [\gamma_p (j), K_0(z,\r)] = r_p z^{jh} K_0(z,\r) .$$

Consider the following operators on the Fock space $F$:
$$L_\tau (j) = {1\over 2} \sum_{i\in\Z} {\bf :}\tau(1-i) \tau(i+j\ell){\bf :} \hbox{\ \ },$$
$$L_{\gamma\kappa} (j) = \sum_{p=1}^n \sum_{i\in\Z}
{\bf :}\gamma_p (-i) \kappa_p (i+j){\bf :} 
\hbox{\ \ } .$$
These are the analogues of the Sugawara operators.

Construct the formal generating series for these:
$$ D_s (z) = - \sum_{j\in\Z} \bigl( L_\tau (j) + h L_{\gamma\kappa} (j) \bigr)
z^{-jh} .$$

{\bf Proposition 8.} 
{\it Formula (3.5) together with }
$$ \sum_{i\in\Z} \phi (s^{ih} \t^\r D_s) z^{-ih} = 
D_s (z, \r) =  {\bf :}D_s (z) K_0 (z,\r){\bf :} $$
{\it defines the representation of the Lie algebra $\D$
on the space}
$$\phi (\hat\g) =  
\phi( \sum\limits_{j\in\Z} \dg_j \otimes s^j \C[t_1^\pm,\ldots, t_n^\pm]  
\oplus \K ).$$

{\it Proof.}
 We have an action of $\D$ on $\hat\g$ which is a unique extension
of its natural action on $\tg$. Thus we need to prove that 
$\phi([D,B]) = [\phi(D), \phi(B)]$ for every $D\in\D$ and $B\in\hat\g$. 
For subalgebra $\D_+$ this was proved in the course of Theorem 4, hence
we need to consider only $D = s^{ih} \t^\r D_s$. Since $s^{ih} \t^\r D_s$
act on $\hat\g$ as derivations and $\hat\g$ is generated as an algebra
by elements $A^\alpha_j \otimes s^j \t^\m$ then it is sufficient to prove that
$$\phi( [ s^{ih} \t^\r D_s, A^\alpha_j \otimes s^j \t^\m ] ) =
[ \phi( s^{ih} \t^\r D_s) , \phi( A^\alpha_j \otimes s^j \t^\m ) ] .$$
Again we shall replace this with an equivalent identity for 
the corresponding series:
$$ \phi( [\sum_{i\in\Z} s^{ih} \t^\r D_s z_1^{-ih} , 
\sum_{j\in\Z} A^\alpha_j \otimes s^j \t^\m z_2^{-j} ]) = 
[ \sum_{i\in\Z} \phi( s^{ih} \t^\r D_s ) z_1^{-ih} , 
\sum_{j\in\Z} \phi( A^\alpha_j \otimes s^j \t^\m ) z_2^{-j} ] .\eqno{(5.1)}$$
   
We have
$$ \phi( [\sum_{i\in\Z} s^{ih} \t^\r D_s z_1^{-ih} , 
\sum_{j\in\Z} A^\alpha_j \otimes s^j \t^\m z_2^{-j} ]) = \hbox{\hskip 5cm}$$
$$\eqalign{
&= \sum_{i\in\Z} \sum_{j\in\Z} j \phi( A^\alpha_j 
\otimes s^{j+ih} \t^{\r+\m} ) z_1^{-ih} z_2^{-j} =
\cr
&= \sum_{i\in\Z} \sum_{k=j+ih\in\Z} (k-ih) 
\phi( A^\alpha_k \otimes s^{k} \t^{\r+\m} ) z_1^{-ih} z_2^{-k+ih} =
\cr
&= \sum_{k\in\Z} k \phi( A^\alpha_k \otimes s^{k} \t^{\r+\m} ) z_2^{-k}
\delta (\zba^h)  - 
\cr
&\quad
h \sum_{k\in\Z} \phi( A^\alpha_k \otimes s^{k} \t^{\r+\m} ) z_2^{-k}
D\delta (\zba^h) =
\cr
&= - \bigl( D_{z_2} A^\alpha(z_2, \r+\m) \bigr) \delta( \zba^h ) -
\cr
&\quad
A^\alpha(z_2, \r+\m) \bigl( D_{z_2} \delta( \zba^h ) \bigr) =
\cr} $$
$$= - D_{z_2} \bigl( A^\alpha(z_2, \r+\m) \delta( \zba^h ) \bigr) ,$$
where $D_{z_2} = z_2 {\d\over\d z_2}$.

In order to compute the right hand side of (5.1), we will need
two lemmas.

{\bf Lemma 9.} 
$$ [L_\tau (j), A^\alpha(z) ] = z^{jh} (D_z + jh) A^\alpha (z) ,
\leqno{(i)} $$
$$ [L_\tau (z_1), A^\alpha(z_2) ] = 
D_{z_2} \bigl( A^\alpha(z_2) \delta( \zba^h ) \bigr) . \leqno{(ii)} $$

{\it Proof.}
 We will prove (i) assuming that $j\geq 0$. The case $j<0$ is
analogous. 
$$ [L_\tau (j), A^\alpha(z) ] = 
{1\over 2} [ \sum_{i\in\Z} {\bf :}\tau (1-i) \tau (i+j\ell){\bf :} ,
 A^\alpha (z) ] =$$
$$ = {1\over 2} [ \sum_{i>0} \tau (1-i) \tau (i+j\ell) , A^\alpha (z) ] +
{1\over 2} [ \sum_{i\leq 0}  \tau (i+j\ell) \tau (1-i) , A^\alpha (z) ] =$$
$$ = {1\over 2}  \sum_{i>0} \tau (1-i) [ \tau (i+j\ell) , A^\alpha (z) ] +
{1\over 2}  \sum_{i>0} [ \tau (1-i) , A^\alpha (z) ] \tau (i+j\ell) +$$
$${1\over 2}  \sum_{i\leq 0}  \tau (i+j\ell) [ \tau (1-i) , A^\alpha (z) ] +
{1\over 2}  \sum_{i\leq 0}  [ \tau (i+j\ell) , A^\alpha (z) ] \tau (1-i)  =$$
$$ = {1\over 2}  \sum_{i>0} \tau (1-i) \lambda^\alpha_i z^{b_i + jh} 
A^\alpha (z) +
{1\over 2}  \sum_{i>0} \lambda^\alpha_{1-i} z^{-b_i} A^\alpha (z) 
\tau (i+j\ell) +$$
$${1\over 2}  \sum_{i\leq 0}  \tau (i+j\ell)  \lambda^\alpha_{1-i} z^{-b_i} 
A^\alpha (z) +
{1\over 2}  \sum_{i\leq 0} \lambda^\alpha_i z^{b_i+jh} A^\alpha (z) 
\tau (1-i)  =$$
$$ = {1\over 2}  \sum_{i>0} \tau (1-i) \lambda^\alpha_i z^{b_i + jh} 
A^\alpha (z) +
{1\over 2}  \sum_{i>0} \lambda^\alpha_{1-i} z^{-b_i} A^\alpha (z) 
\tau (i+j\ell) +$$
$${1\over 2}  \sum_{i\leq -j\ell}  \tau (i+j\ell)  \lambda^\alpha_{1-i} 
z^{-b_i} A^\alpha (z) +
{1\over 2}  \sum_{-j\ell<i\leq 0} \lambda^\alpha_{1-i} z^{-b_i} 
[\tau (i+j\ell) , A^\alpha (z) ] +$$
$${1\over 2}  \sum_{-j\ell<i\leq 0}  \lambda^\alpha_{1-i} z^{-b_i} 
A^\alpha (z) \tau (i+j\ell) +
{1\over 2}  \sum_{i\leq 0} \lambda^\alpha_i z^{b_i+jh} A^\alpha (z) 
\tau (1-i)  =$$
$$ = {1\over 2}  \sum_{k=i>0} \tau (1-k) \lambda^\alpha_k z^{b_k + jh} 
A^\alpha (z) +
{1\over 2}  \sum_{k=i+j\ell>j\ell} \lambda^\alpha_{1-k} z^{-b_k+jh} 
A^\alpha (z) \tau (k) +$$
$${1\over 2}  \sum_{k=1-i-j\ell>0} \lambda^\alpha_{k} z^{b_k + jh}
 \tau (1-k) A^\alpha (z) +
{1\over 2}  \sum_{-j\ell<i\leq 0} \lambda^\alpha_{1-i} z^{-b_i} 
\lambda^\alpha_i z^{b_i+jh} A^\alpha (z) +$$
$${1\over 2}  \sum_{0<k=i+j\ell \leq j\ell}  \lambda^\alpha_{1-k} 
z^{-b_k +jh} 
A^\alpha (z) \tau (k) +
{1\over 2}  \sum_{k=1-i>0} \lambda^\alpha_{1-k} z^{-b_k+jh} A^\alpha (z) 
\tau (k) =$$
$$ = z^{jh} D_z A^\alpha (z) + {j\over 2} z^{jh} 
\bigl( \sum_{k=1}^\ell \lambda^\alpha_k \lambda^\alpha_{1-k} \bigr) 
A^\alpha (z) =$$
$$ = z^{jh} ( D_z  + hj ) A^\alpha (z) .$$

At the last step we used the equality
$${1\over 2} \sum_{k=1}^\ell \lambda^\alpha_k \lambda^\alpha_{1-k} =
{1\over 2} \sum_{k=1}^\ell (T_k | \nu^{-1} (\alpha) ) (T_{1-k} | \nu^{-1}(\alpha))
=$$
$${h\over 2} (\nu^{-1}(\alpha) | \nu^{-1}(\alpha) ) =
  {h\over 2} (\alpha | \alpha) = h ,$$
which follows from the fact that
$\{ {1\over h} T_{1-k} \}$ is the dual basis for $\{ T_k \}$.

Part (ii) is an immediate consequence of (i):
$$ [L_\tau (z_1), A^\alpha(z_2) ] = 
\sum_{j\in\Z} z_1^{-jh} [L_\tau (j), A^\alpha(z_2)]  =$$
$$ \sum_{j\in\Z} z_1^{-jh} z_2^{jh} (D_{z_2} + hj) A^\alpha (z_2) =
\bigl( D_{z_2} A^\alpha (z_2) \bigr) \delta (\zba^h) +
A^\alpha (z_2) \bigl( D_{z_2} \delta (\zba^h) \bigr) =$$
$$D_{z_2} \bigl( A^\alpha(z_2) \delta( \zba^h ) \bigr) .$$

{\bf Lemma 10.}
$$ [ {\bf :}L_{\gamma\kappa} (z_1) K_0 (z_1,\r){\bf :} , K_0 (z_2, \m) ] =
{1\over h} \delta( \zba^h ) D_{z_2} (K_0(z_2,\m)) K_0 (z_1,\r) .$$

{\it Proof.}
$$ [ {\bf :}L_{\gamma\kappa} (z_1) K_0 (z_1,\r){\bf :} , K_0 (z_2, \m) ] =$$
$$= \sum_{j\in\Z} z_1^{-jh} \sum_{p=1}^n \sum_{i>0}
[\gamma_p (-i) \kappa_p (i+j) K_0 (z_1,\r) , K_0 (z_2, \m) ] +$$
$$\sum_{j\in\Z} z_1^{-jh} \sum_{p=1}^n \sum_{i\leq 0}
[\kappa_p (i+j) K_0 (z_1,\r) \gamma_p (-i) , K_0 (z_2, \m) ] =$$
$$= \sum_{j\in\Z} z_1^{-jh} \sum_{p=1}^n \sum_{i>0}
[\gamma_p (-i) , K_0 (z_2, \m) ] \kappa_p (i+j) K_0 (z_1,\r) +$$
$$\sum_{j\in\Z} z_1^{-jh} \sum_{p=1}^n \sum_{i\leq 0}
\kappa_p (i+j) K_0 (z_1,\r) [ \gamma_p (-i) , K_0 (z_2, \m) ] =$$
$$= \sum_{j\in\Z} z_1^{-jh} \sum_{p=1}^n m_p \sum_{i>0}
z_2^{-ih} K_0 (z_2, \m) \kappa_p (i+j) K_0 (z_1,\r) +$$
$$\sum_{j\in\Z} z_1^{-jh} \sum_{p=1}^n m_p \sum_{i\leq 0}
\kappa_p (i+j) K_0 (z_1,\r) z_2^{-ih} K_0 (z_2, \m) =$$
$$= \sum_{j\in\Z} \sum_{k=i+j\in\Z} \sum_{p=1}^n m_p 
z_1^{-jh} z_2^{-kh+jh} \kappa_p (k) K_0 (z_2, \m) K_0 (z_1,\r) =$$
$$ = {1\over h} \delta( \zba^h ) D_{z_2} (K_0 (z_2, \m)) K_0 (z_1, \r) .$$       

\

Using these two lemmas we can handle the right hand side of (5.1). 
$$[ \sum_{i\in\Z} \phi( s^{ih} \t^\r D_s ) z_1^{-ih} , 
\sum_{j\in\Z} \phi( A^\alpha_j \otimes s^j \t^\m ) z_2^{-j} ] =
[ D_s (z_1, \r) , A^\alpha (z_2, \m) ] = $$
$$\eqalign{
&= [ {\bf :}D_s (z_1) K_0 (z_1, \r){\bf :} , A^\alpha (z_2) K_0 (z_2, \m) ] = 
\cr
&= - [ L_\tau (z_1) K_0 (z_1, \r) , A^\alpha (z_2) K_0 (z_2, \m) ] - 
\cr
&\quad
- h [ {\bf :}L_{\gamma\kappa} (z_1) K_0 (z_1, \r){\bf :} , 
A^\alpha (z_2) K_0 (z_2, \m) ] = 
\cr
&\quad
 - [ L_\tau (z_1) , A^\alpha (z_2) ] K_0 (z_1, \r) K_0 (z_2, \m) - 
\cr
&\quad
h A^\alpha (z_2) 
[ {\bf :}L_{\gamma\kappa} (z_1) K_0 (z_1, \r){\bf :} , K_0 (z_2, \m) ] =  
\cr
&= - D_{z_2} \bigl( A^\alpha(z_2) \delta( \zba^h ) \bigr) K_0 (z_2, \m) 
K_0 (z_1, \r) - 
\cr
&\quad
A^\alpha (z_2) \delta( \zba^h ) D_{z_2} (K_0 (z_2, \m)) K_0 (z_1, \r) = 
\cr}$$
$$\eqalign{ 
&= - D_{z_2} \bigl( A^\alpha(z_2) \delta( \zba^h ) K_0 (z_2, \m) K_0 (z_1, \r)
\bigr) = 
\cr
&=  - D_{z_2} \bigl( A^\alpha(z_2) \delta( \zba^h ) K_0 (z_2, \m) K_0 (z_2, \r)
\bigr) = 
\cr
&= - D_{z_2} \bigl( A^\alpha(z_2, \r+\m) \delta( \zba^h ) \bigr) .
\cr}$$

\vfill\eject

{\bf References}

\

\item{[BM]} G.M.~Benkart, R.V.~Moody, {\it Derivations,
central extensions, and affine Lie algebras,}
Algebras Groups Geom., {\bf 3} (1986), 456-492.

\item{[BC]}  S.~Berman, B.~Cox, {\it Enveloping algebras and representations
of toroidal Lie algebras,} Pacific J.Math., {\bf 165} (1994), 239-267.

\item{[Co1]} H.S.M.~Coxeter, {\it Regular polytopes,}
Methuen, London, 1948.

\item{[Co2]} H.S.M.~Coxeter, {\it Regular complex 
polytopes,}
Cambridge University Press, 1974.

\item{[EM]} S.~Eswara Rao, R.V.~Moody, {\it Vertex representations for $n$-toroidal
Lie algebras and a generalization of the Virasoro algebra,} 
Comm. Math. Phys. {\bf 159} (1994), 239-264.

\item{[FLM]} I.~Frenkel, J.~Lepowsky, A.~Meurman, {\it Vertex operator 
algebras and the Monster,} Academic Press, Boston, 1989.

\item{[IKUX]} T.~Inami, H.~Kanno, T.~Ueno, C.-S.~Xiong,
{\it Two-toroidal Lie algebra as current algebra
of four-dimensional K\"ahler WZW model,}
hepth/9610187. 

\item{[Kac]} V.G.~Kac, {\it Infinite-dimensional Lie algebras,} 3rd ed.,
Cambridge University Press, 1990.

\item{[KKLW]} V.G.~Kac, D.A.~Kazhdan, J.~Lepowsky, R.L.~Wilson, {\it
Realization of the basic representation of the Euclidean Lie algebras,}
Adv.Math., {\bf 42} (1981), 83-112.

\item{[Kas]} C.~Kassel, {\it K\"ahler differentials and coverings of complex
simple Lie algebras extended over a commutative algebra,}
J.Pure Appl.Algebra {\bf 34} (1985), 265-275.

\item{[Kos]} B.~Kostant, {\it The principal 
three-dimensional subgroup and the Betti numbers of 
a complex simple Lie group,} Amer.J.Math. {\bf 81}
(1959), 973-1032.

\item{[MEY]} R.V.~Moody, S.~Eswara Rao, T.~Yokonuma, {\it Toroidal Lie algebras 
and vertex representations,} Geom.Ded., {\bf 35} (1990), 283-307.

\end